\documentclass[12pt,preprint]{aastex}

\newcommand{\Msolar}{\mbox{\,M$_\odot$}}        
\newcommand{\kms}{\mbox{\,km\,s$^{-1}$}}                           
\newcommand{\degrees}{$^{\circ}$}

\newcommand{\um}{$\mu$m}
\newcommand{\xco}{$^{13}$CO}
\newcommand{\co}{$^{12}$CO}

\slugcomment{V3. This draft \today. \\
For resubmission to The Astronomical Journal.}
\shorttitle{COMPLETE Phase I Data}
\shortauthors{Ridge et al.}

\begin{document}
\title{The COMPLETE Survey of Star Forming Regions: Phase I Data}

\author{
Naomi A. Ridge\altaffilmark{1},
James Di Francesco\altaffilmark{2},
Helen Kirk\altaffilmark{2,3},
Di Li\altaffilmark{1,4},
Alyssa A. Goodman\altaffilmark{1},
Jo\~ao F. Alves\altaffilmark{5},
H\'ector G. Arce\altaffilmark{6},
Michelle A. Borkin\altaffilmark{7},
Paola Caselli\altaffilmark{8,1},
Jonathan B. Foster\altaffilmark{1},
Mark H. Heyer\altaffilmark{9},
Doug Johnstone\altaffilmark{2,3},
David A. Kosslyn\altaffilmark{1},
Marco Lombardi\altaffilmark{4}
Jaime E. Pineda\altaffilmark{1},
Scott L. Schnee\altaffilmark{1},
Mario Tafalla\altaffilmark{10}
}

\altaffiltext{1}{Harvard-Smithsonian Center for Astrophysics, 60 Garden Street, Cambridge, MA, 02138, USA}
\altaffiltext{2}{National Research Council of Canada, Herzberg Institute of Astrophysics, 5071 West Saanich Road, Victoria, BC, V9E 2E7, Canada}
\altaffiltext{3}{Department of Physics and Astronomy, University of Victoria, P.O. Box 3055, Station CSC, Victoria, BC, V8P 1A1, Canada}
\altaffiltext{4}{JPL/Caltech, 4800 Oak Grove Drive, Pasadena, CA 91109, USA}
\altaffiltext{5}{European Southern Observatory, Karl-Schwartzschild-Strasse 2, D-85748 Garching bei M\"unchen, Germany}
\altaffiltext{6}{Department of Astrophysics, American Museum of Natural History, New York, NY, 10024, USA}
\altaffiltext{7}{Dept. of Astronomy, Harvard University, 60 Garden Street, Cambridge, MA, 02138, USA }
\altaffiltext{8}{INAF - Osservatorio Astrofisico di Arcetri, Largo E. Fermi 5, 50125 Firenze, Italy}
\altaffiltext{9}{Department of Astronomy, University of Massachusetts, Lederle Graduate Research Center, Amherst, MA, 01003, USA}
\altaffiltext{10}{Observatorio Astron\'omico Nacional (IGN), Alfonso XII, 3, E-28014 Madrid, Spain}

\begin{abstract}
We present an overview of data available for the Ophiuchus and Perseus
molecular clouds from ``Phase I'' of the COMPLETE Survey of
Star-Forming Regions. This survey provides a range of data
complementary to the Spitzer Legacy Program ``From Molecular Cores to
Planet Forming Disks.''  Phase I includes: Extinction maps derived
from 2MASS near-infrared data using the NICER algorithm; extinction
and temperature maps derived from IRAS 60 and 100\um\ emission; HI
maps of atomic gas; \co\ and \xco\ maps of molecular gas; and
submillimetre continuum images of emission from dust in dense cores.
Not unexpectedly, the morphology of the regions appears quite
different depending on the column-density tracer which is used, with
IRAS tracing mainly warmer dust and CO being biased by chemical,
excitation and optical depth effects.  Histograms of column-density
distribution are presented, showing that extinction as derived from
2MASS/NICER gives the closest match to a log-normal distribution as is
predicted by numerical simulations.  All the data presented in this
paper, and links to more detailed publications on their implications
are publically available at the COMPLETE website.

\end{abstract}


\section{Introduction}
The prevailing theoretical picture of star formation envisions stars
forming within dense cores, which are embedded in turn within larger,
slightly lower-density structures.  Each forming star is surrounded by
a disk, and, when it is very young, the star-disk system produces a
collimated bipolar flow, in a direction perpendicular to the disk.  In
its broad outlines, this paradigm is very likely to be right.  In
detail, however, many questions remain concerning the timing of this
series of events.  For example, how long does a star stay with its
natal core?  How long does it remain associated with the lower-density
structure (e.g., a filament in a dark cloud) where it originally
formed?  What kind of environment does a star-disk system need to keep
accreting or to produce an outflow -- and when might that reservoir no
longer be available to the system?  Does a bipolar outflow have any
effect on star formation nearby?  What causes fragmentation into
binaries or higher-order systems?  How much influence do spherical
winds (e.g., SNe, B-star winds) from previous generations of stars
have on the timing of star formation?  How often is a star in the
process of forming likely to encounter an external gravitational
potential (e.g., from another forming star) strong enough to alter its
formation process?  The complicating issue underlying all of these
questions is that it is hard to define and understand the
properties of the reservoir from which a star forms if the reservoir
itself is highly dynamic.

The COMPLETE\footnote{http://cfa-www.harvard.edu/COMPLETE} ({\bf
C}o{\bf O}rdinated {\bf M}olecular {\bf P}robe {\bf L}ine, {\bf
E}xtinction and {\bf T}hermal {\bf E}mission) Survey of Star-Forming
Regions is intended to provide an unprecedented comprehensive database
with which one might have real hope of answering many statistically
addressable questions about star formation.  Its primary goal is to
provide detailed measurements of the velocity fields (from molecular
line observations), the density profiles (from extinction
measurements), the temperature and dust property profiles (from
thermal emission mapping), the larger cloud environment (from atomic
hydrogen) and the embedded source distributions (from infrared
imaging) of several nearby molecular clouds.

COMPLETE is not the first such study of this kind -- \citet{lada92}
mapped the actively star-forming cloud \objectname{L1630} (a.k.a. 
\objectname{Orion B}) in both molecular line emission 
(CS, with 2$'$ resolution) 
and with near-infrared cameras (reaching $m_K<13$ mag).  That work
provided the first evidence that massive stars form in clusters, and
it also showed that the mass spectrum of the gaseous material
(self-gravitating or not) is shallower than that for stars.  COMPLETE,
which was not feasible a decade ago, is intended to allow for low-mass
(fainter) star-forming regions what \citeauthor{lada92}'s work allowed
for (brighter) massive star-forming regions, and more.  For reference,
the total areal coverage of COMPLETE is $\sim 20$ square degrees,
which is an order of magnitude larger than the area
\citeauthor{lada92} studied in Orion.

COMPLETE makes it possible for researchers to combine diverse
observational techniques and to measure the physical properties of the
three star-forming clouds, Perseus, Ophiuchus and Serpens. These three
were chosen because they are nearby extended cloud targets and were
also included in the Spitzer Space Telescope ``From Molecular Cores to
Planet Forming Disks" (c2d) Legacy Program
\citep{evans03} that are visible from northern hemisphere
observatories.  Selecting such targets from c2d ensured that full
censes of their embedded stellar populations and their properties
would be available around the same time that COMPLETE was finished.

COMPLETE was designed to be executed in two phases.  Phase I focuses
on observing the ``larger context'' of the star formation process (on
0.1 to 10 pc scales).  {\it All} of the Perseus and Ophiuchus cloud
area to be observed with Spitzer under c2d, and slightly more, have
been already covered by COMPLETE in an unbiased way for Phase I.
Observations of the third cloud, Serpens, are less advanced and hence
are not included here\footnote{Serpens data are available at the
COMPLETE website however.}.  Phase $\amalg$, which is already well underway,
will be a statistical study of the small-scale picture of star
formation, aimed at assessing the meaning of the variety of physical
conditions observed in star forming cores (on the $<0.1$ pc scale).
In Phase $\amalg$, targeted source lists based on the Phase I data are being
used, as it is (still) not feasible to cover {\it every} dense
star-forming peak at high resolution. These data are being released on
the COMPLETE website as they are validated.

In this paper, we present an overview of the Phase I data of COMPLETE
for the Ophiuchus and Perseus clouds: extinction maps derived from
near-infrared data; extinction and temperature maps derived from IRAS
60 and 100\,\um\ emission; emission maps of atomic line data (HI);
emission maps of molecular line data (\co\ and \xco); and emission
maps of submillimetre continuum data.  Section \ref{obs} gives a
description of the data acquistion and reduction techniques for each
data set. Detailed analyses and interpretation of these data appear
elsewhere \citep[e.g., ][]{johnstone04, wala05, schnee_iras,
ridge_iras, scott_bethell, gsr, ridge_co, helen}. All Phase I COMPLETE
data are publically available and can be retrieved from the COMPLETE
website, http://cfa-www.harvard.edu/COMPLETE.

\section{Data and Observational Findings}
\label{obs}

Figures \ref{oph_coverage} and \ref{per_coverage} show the 2MASS/NICER
extinction maps for Ophiuchus and Perseus (see section \ref{nicer})
overlaid with the boundaries of the associated regions surveyed in
molecular lines, thermal dust emission and HI emission, as described
in the remainder of this section.  COMPLETE coverage was chosen to
coincide with the coverage of c2d IRAC observations, and to include
most areas with A$_V > 3$ (as indicated by the white contour in
figures \ref{oph_coverage} and \ref{per_coverage}) and all with A$_V >
5$.  Table \ref{props_tab} summarises the basic physical properties of
the two star-forming regions.

\begin{deluxetable}{ccccc}
\tablecolumns{5}
\tablewidth{0pc}
\tablecaption{Ophiuchus and Perseus physical properties
\label{props_tab}}
\tablehead{
\colhead{}&
\colhead{Distance\tablenotemark{1} (pc)}&
\colhead{Total Gas Mass\tablenotemark{2} (\Msolar)}&
\colhead{Size\tablenotemark{3} (pc$^2$)}&
\colhead{Refs.}
}
\startdata
Ophiuchus & 125$\pm$25&7.4$\times 10^3$&46&\citet{dG89}\\
Perseus & 250$\pm$50&1.0$\times 10^4$&70&\citet{enoch}\\
\enddata
\tablenotetext{1}{We quote the distances adopted by the c2d team
for each of the clouds (N. Evans, personal communication).}
\tablenotetext{2}{Total mass enclosed within the A$_V$=3 contour
as determined from the 2MASS/NICER extinction map.}
\tablenotetext{3}{Total area enclosed within the A$_V$=3 contour as determined from the 2MASS/NICER extinction map.}
\end{deluxetable}

\clearpage

\begin{figure}
\caption{
2MASS/NICER extinction map of Ophiuchus overlaid with outlines showing
the areas covered by \co\ and \xco\ observations (green), 850\,\um\
continnum observations (red) and HI observations (yellow) which are
all available to download from the COMPLETE website.  Note that the
small ``hole'' at the center of the L1689 cluster is an artifact due
to the high extinction at that position. The white contour indicates
an A$_V$ of 3\,mag.
\label{oph_coverage}}
\end{figure}

\clearpage

\begin{figure}
\caption{2MASS/NICER extinction map of Perseus overlaid with outlines showing the areas covered by \co\ and \xco\ observations (green), 850\um\ continnum observations (red) and HI observations (yellow) which are
all available to download from the COMPLETE website. The white contour
indicates an A$_V$ of 3\,mag. The black and white crosses indicate the
positions of the spectra shown in figures \ref{b1_h1} and \ref{wing}
respectively.
\label{per_coverage}}
\end{figure}

\clearpage

\subsection{Extinction Maps from 2MASS}
\label{nicer}

Near infrared extinction maps for Ophiuchus and Perseus were produced
from the final data release of the point source catalog from the Two
Micron All-Sky Survey\footnote{The 2MASS project is a collaboration
between The University of Massachusetts and the Infrared Processing
and Analysis Center (JPL/Caltech). Funding is provided primarily by
NASA and the NSF.} (2MASS; see \\
http://www.ipac.caltech.edu/2mass/releases/allsky/doc/explsup.html).
We used the NICER (Near-Infrared Color Excess Revisited) algorithm
\citep{la01,laf04}, which combines observations from all three
near-infrared bands to produce maps with lower noise than is possible
from using just two bands.  The NICER algorithm takes advantage of the
small variation in intrinsic colors of stars in the near-infrared to
obtain an accurate estimate of the column density towards each star.
In particular, the colors of all 2MASS stars in the field are compared
to the colors of (supposedly) unreddened stars in a control field;
then, for each star, an optimal combination of near-infrared colors is
determined by taking into account the different response of the
various colors to the reddening, the photometric errors on the star
magnitudes, and the dispersion of intrinsic colors (as determined from
the control field; see \citet{la01} for further details).  Note that
for the NICER analysis we used a normal reddening law
\citep{rl85}; interestingly, this reddening law
compares well with the newly determined 2MASS reddening law
(\citealt[][; see also \citealt{lal06}]{indeb05}).

At the end of this preliminary step, we have a catalog of extinction
measurements (and relative errors) for each star.  These pencil-beam
estimates need then to be interpolated in order to produce a smooth
map and to reduce their variance.  In particular, we smoothed the data
using the moving average technique using a Gaussian kernel.  This
simple smoothing algorithm has easily quantified error properties
(e.g., it is possible to easily determine the error map, see
http://cfa-www.harvard.edu/COMPLETE/data\_html\_pages/2MASS.html), and
works well for nearby objects such as those in COMPLETE where
background stars constitute the vast majority of all stars, and do not
force us to either clip out high-sigma (foreground) stars or employ
the more robust weighted median, both of which would lead to more
complicated error properties.  Smoothed in this way, the map is really
the convolution of the true extinction with the weighting function,
and so it is not sensitive to spatial scales smaller than the
weighting function \citep[see ][]{ls01}.

In Perseus, we produced an extinction map which is approximately
9\degrees\ by 12\degrees, with an effective resolution of 5$'$.  The
average 1-$\sigma$ noise in this image is 0.18 magnitudes of $A_V$.
Some remnant stripes are visible in the north-south direction of the
2MASS strips, reflecting errors in 2MASS calibration between strips.
In Ophiuchus, our map is 9\degrees\ by 8\degrees, with an effective
resolution of 3$'$, and an average 1$\sigma$ noise of 0.16 magnitudes
of $A_V$.  Again, 2MASS stripes are visible.  The properties of the
maps are summarised in table \ref{2mass_data}.
\begin{deluxetable}{ccc}
\tablewidth{0pc}
\tablecaption{2MASS/NICER Data
\label{2mass_data}}
\tablehead{\colhead{}&\colhead{Ophiuchus}&\colhead{Perseus}}
\startdata
Pixel Size ($'$) & 1.5 & 2.5\\
Effective Resolution ($'$) &3 & 5\\
Areal Coverage (degrees)& 9$\times$8 & 9$\times$12\\
1$\sigma$ noise (A$_V$) & 0.16 & 0.18\\
\enddata
\end{deluxetable}

A small hole is present in the center of L1688 in the Ophiuchus map,
where 2MASS provides no data due to extremely high
extinction\footnote{Deeper near-infrared observations being obtained
as part of Phase $\amalg$ will fill such holes.} (i.e., $A_V > 30$).  

Several globular clusters, listed in table \ref{globs} also show up in
regions of relatively little extinction. Stars in these clusters are
typically rather blue, and thus produce slightly lower extinction
values. The error map is useful for identifying these globular
clusters, as well as star-forming clusters associated with either
cloud, both of which bias the extinction determination but show up in
the error map as regions of very low error since the stellar density
is high \citep{laf04}.
\begin{deluxetable}{ccc}
\tablewidth{0pc}
\tablecaption{Globular Clusters which affect the Ophiuchus extinction map
\label{globs}}
\tablehead{\colhead{Name}&\colhead{RA\tablenotemark{a}}&\colhead{Dec\tablenotemark{b}}}
\startdata
NGC 6235& 16:53:25.36& -22 10 38.8\\
NGC 6144& 16 27 14.14&  -26 01 29.0\\
M80& 16 17 02.51&  -22 58 30.4\\
M4& 16 23 35.41&  -26 31 31.9
\enddata
\tablenotetext{a}{J2000. Units are hh:mm:ss.ss}
\tablenotetext{b}{J2000. Units are dd:mm:ss.s}
\end{deluxetable}

The final extinction maps, with the locations of the well-known dense
cores and star-forming clusters in the two regions indicated, are
shown in figures \ref{oph_2mass} and \ref{per_2mass}.

The map of Ophiuchus reveals well its multi-filamentary structure;
starting from the very opaque L1688\footnote{The bright extinction
peak we have labelled ``L1688'' also includes L1686, L1692, L1690 and
L1681} to the west, two filaments can be seen, a northeast filament
(including L1740) and a less tenuous east filament, containing the
L1729, L1712 and L1689 Lynds clouds. The
northeast filament appears the longest and has extinction maxima at
each end associated with L1765 and L1709.
L1688, where the filaments intersect, has the highest extinction in
the Ophiuchus cloud. There is also an extension of L1688 to the
northwest, containing L1687 and L1680.

The 2MASS/NICER extinction map of Perseus (figure \ref{per_2mass})
shows the familiar chain of dark clouds from northeast to
southwest. All of the known dark clouds and star-forming regions are
seen, with the highest extinction regions corresponding to the two
well-studied star-forming clusters IC\,348 and NGC\,1333.

Histograms showing the distribution of extinctions are shown in figure
\ref{2mass_hist}. Both the Ophiuchus and Perseus histograms show an
approximately log-normal distribution of material (indicated by the
grey curve), as is predicted by numerical simulations
(e.g. \citealt{lognormal}). The physical implications of the measured
column-density distributions will be discussed further in
\citet*{gsr}, but note that due to the differing resolutions, 
the histograms presented are not directly comparable to those
presented for the IRAS and CO data in subsequent sections.

\clearpage

\begin{figure}
\caption{Map of extinction in Ophiuchus derived using 2MASS/NICER.
The contour indicates an A$_V$ of 3\,mag. and is repeated in
subsequent figures for orientation. Note that the small ``hole'' at
the center of the L1689 cluster is an artifact due to the high
extinction at that position.
\label{oph_2mass}}
\end{figure}
\begin{figure}
\caption{Map of extinction in Perseus derived using 2MASS/NICER. The contour indicates an A$_V$ of 3\,mag. and is repeated in subsequent figures for orientation. The well-known dark clouds and star-forming clusters in the region are labelled.
\label{per_2mass}}
\end{figure}
\begin{figure}
\plottwo{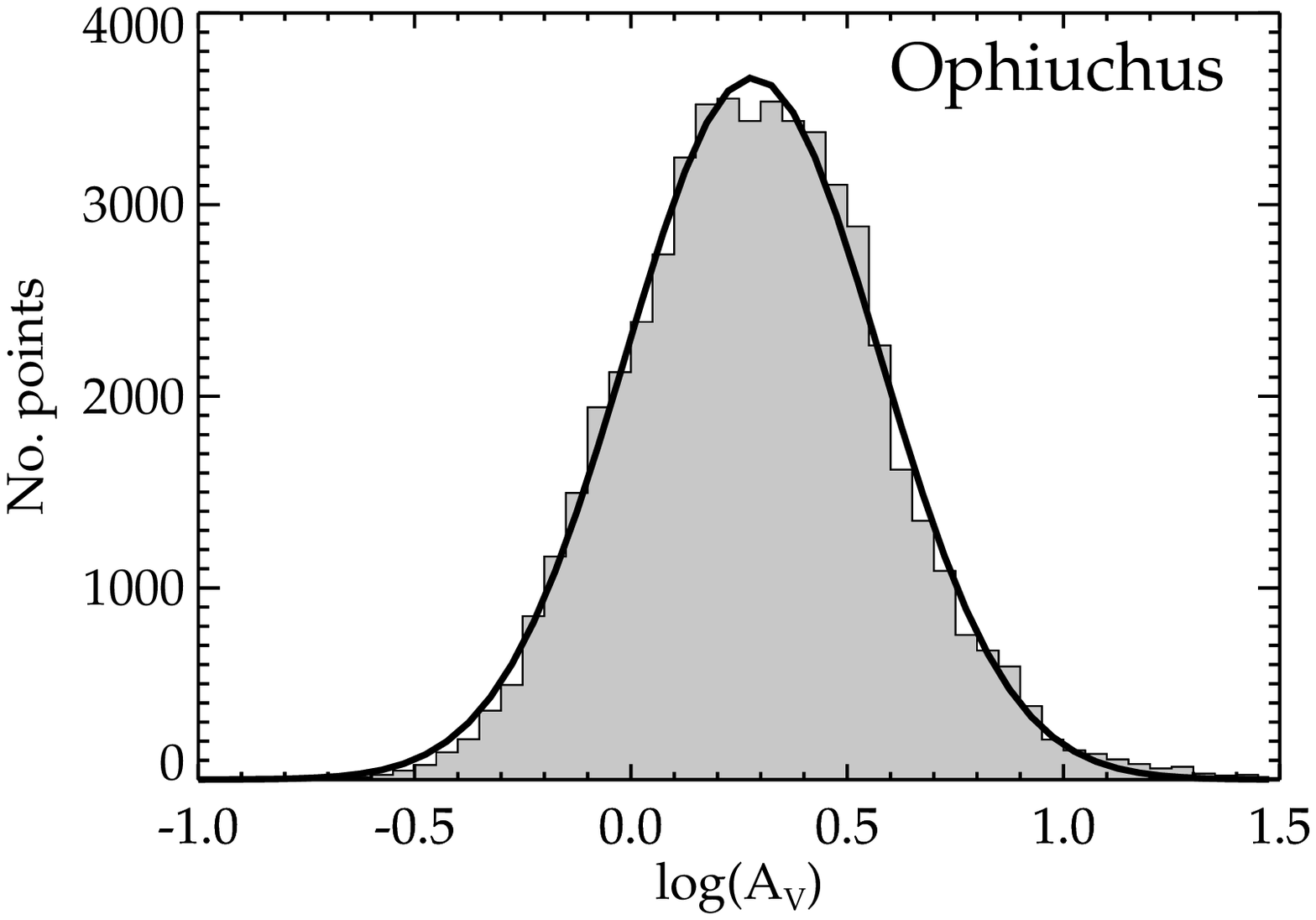}{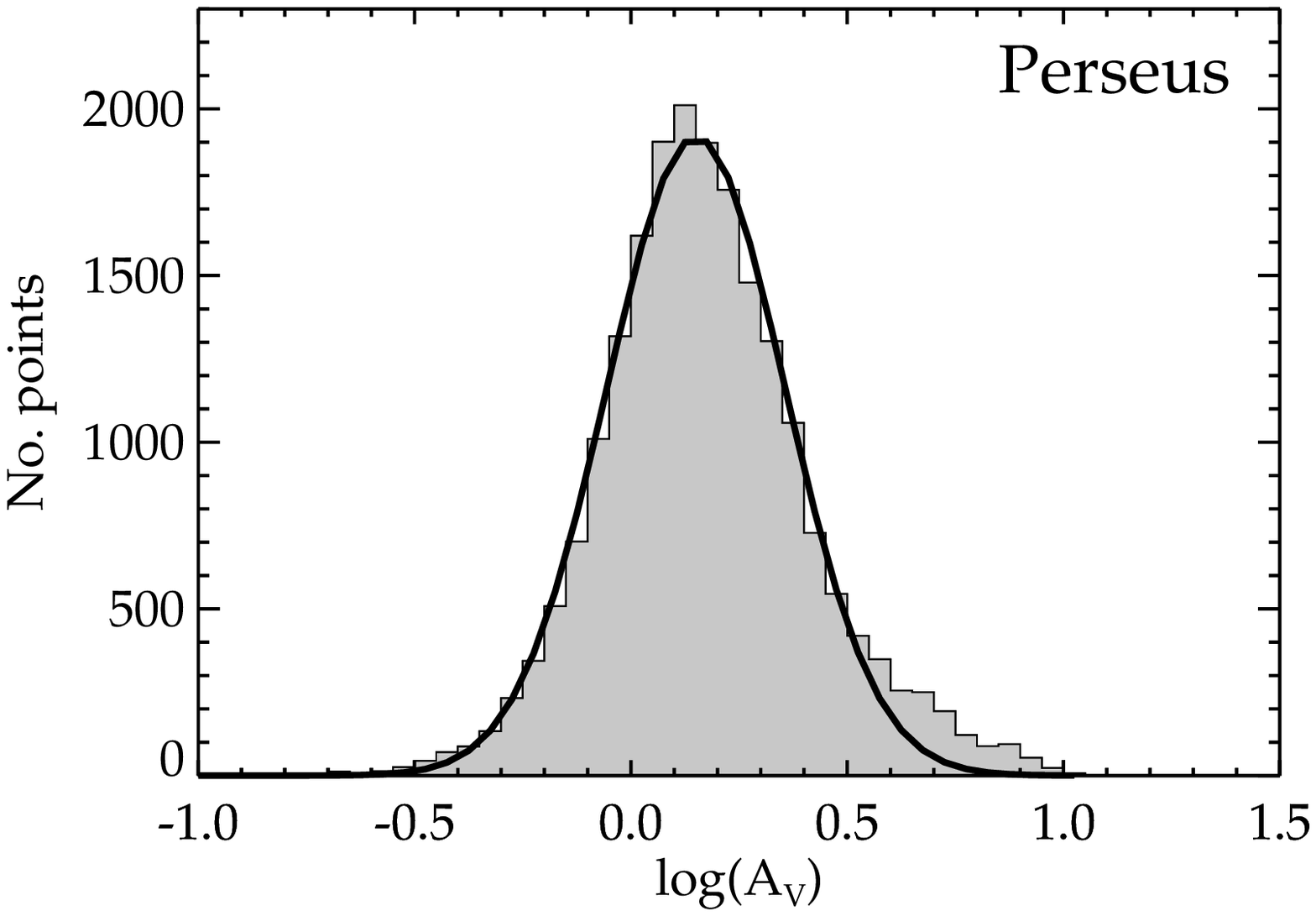}
\caption{Histograms of extinction as derived from 2MASS
in Ophiuchus (left) and Perseus (right) for the areas shown in figures
\ref{oph_2mass} and \ref{per_2mass} (filled histograms) overplotted
with a log-normal fit to the data (thick black curve).
\label{2mass_hist}}
\end{figure}

\clearpage

\subsection{Extinction and Temperature Maps from IRAS}
\label{iras}

IRIS\footnote{Improved Recalibration of the {\it IRAS} Survey} images
\citep{lagache05}
of 60 and 100\,\um\ flux density were obtained for the two
regions. IRIS data offer excellent correction for the effects of
zodiacal dust and striping in the unprocessed IRAS images and also
provide improved gain and offset calibration over earlier releases
(e.g.\ ISSA), which did not have an appropriate zero-point
calibration. This can have serious consequences on the derived dust
temperature and column density \citep{ag99b,ag99a}.

We used the method described in \citet{schnee_iras} (which is
adaptation of the method used by \citealt{wood94} and \citealt{ag99b})
to calculate the dust color temperature and column density from the
IRIS 60 and 100\,\um\ flux densities. The temperature is determined by
the ratio of the 60 and 100\,\um\ flux densities, assuming that the
dust in a single beam is isothermal (the validity and implications of
this assumption are discussed in detail in \citealt{schnee_iras} and
\citealt*{scott_bethell}). Then the column density of dust can be derived
from measured flux and the derived color temperature of the dust. The
calculation of temperature and column density depends on the values of
three parameters: two constants that determine the emissivity spectral
index and the conversion from 100\,\um\ optical depth to visual
extinction. These parameters are solved for explicitly using the
independent estimate of visual extinction we have from our 2MASS/NICER
extinction maps (described in sect. \ref{nicer}).

The 2MASS/NICER extinction maps are used as a ``model'' of the
extinction, and the three free parameters adjusted until the
IRIS-implied column density best matches that of the ``model''.  The
parameter values determined by this method are those that create a
FIR-based extinction map that best matches the 2MASS/NICER extinction
map on a statistical point-by-point basis, and is not a spatial match
to features in the 2MASS/NICER extinction map.

Each cloud is considered separately, so the derived values of the
three parameters are different for each cloud.  We assume that the
values of the three parameters are constant within each image,
although this does not have to be the case. For instance, it is likely
that areas of especially high or low column density do not share the
same far-IR column-density to visual extinction conversion factor.

The IRAS-based extinction and temperature maps for Perseus and
Ophiuchus are shown in figures \ref{oph_iras} and \ref{per_iras} and their 
properties are given in table
\ref{iras_data}.

\begin{deluxetable}{ccc}
\tablewidth{0pc}
\tablecaption{IRAS Data
\label{iras_data}}
\tablehead{\colhead{}&\colhead{Ophiuchus}&\colhead{Perseus}}
\startdata
Pixel Size ($'$) & 5 & 5\\
Effective Resolution ($'$) &5 & 5\\
Areal Coverage (degrees)& 6.3$\times$6.4 & 7$\times$4.5\\
\enddata
\end{deluxetable}

The IRAS-based extinction map of Ophiuchus shows a general similarity
to the 2MASS-based extinction map.  Notable differences, however,
include an extension of moderate extinction to the northwest of L1688
and a regular series of high extinction peaks running west of L1689.
Although there is some enhanced extinction northwest of L1688 in the
2MASS/NICER map, this difference is likely an example of how the IRAS-based
extinction maps can be biased toward warmer dust, and possibly
suggests warmer dust in that region due to external heating by the
star $\rho$ Oph itself
\citep{schnee_iras}. A notable feature in the temperature map is a
heated ring, visible just to the north of L1688 and centered on the
star B-star $\rho$-Oph.

Unlike in Ophiuchus, the IRAS-based extinction map of Perseus (figure
\ref{per_iras}, left panel) shows a vastly different morphology from
the 2MASS/NICER extinction map (on more careful inspection, and by
comparison with the dust color-temperature map (figure \ref{per_iras},
right panel) the known dark cores in Perseus are visible
however). This is because the column density as measured by IRAS is
dominated by a ~0.75\degrees\ warm shell, probably caused by emission
from transiently heated small dust grains at the edge of an H{\sc ii}
region created by the B0 star HD\,278942
\citep{ridge_iras,ander2000}.

\clearpage
\begin{figure}
\caption{Column density (left) and temperature (right)
in Ophiuchus derived from IRAS. The contour indicates A$_V$ of 3\,mag.
as in figure \ref{oph_2mass}.
NOTE TO EDITOR: THIS FIGURE SHOULD BE TYPESET FULL-PAGE, LANDSCAPE ORIENTATION
\label{oph_iras}}
\end{figure}
\begin{figure}
\caption{Column density (left) and temperature (right)
in Perseus derived from IRAS.  The contour indicates A$_V$ of 3\,mag.
as in figure \ref{per_2mass}.
NOTE TO EDITOR: THIS FIGURE SHOULD BE TYPESET FULL-PAGE, LANDSCAPE ORIENTATION
\label{per_iras}}
\end{figure}

\begin{figure}
\plottwo{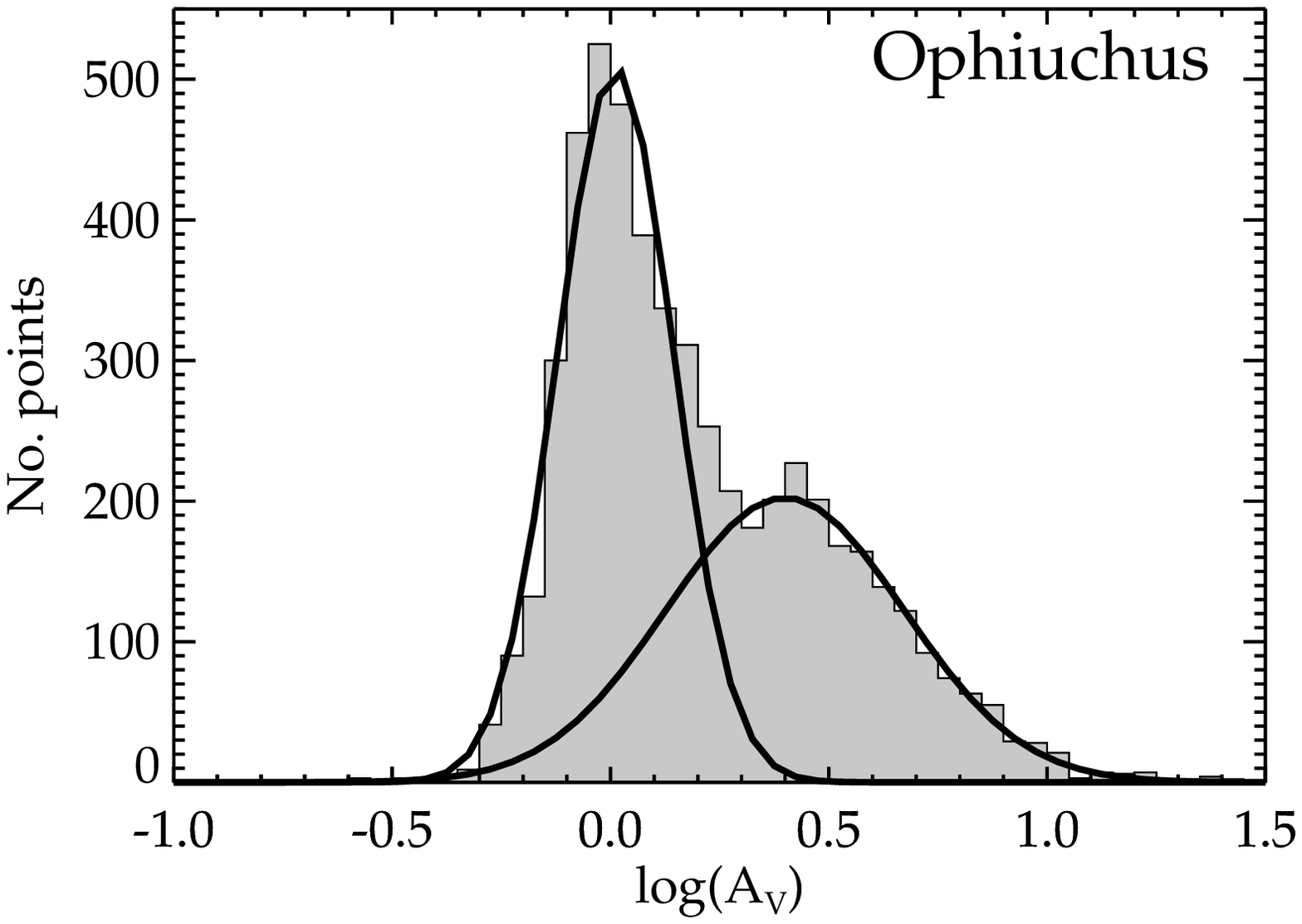}{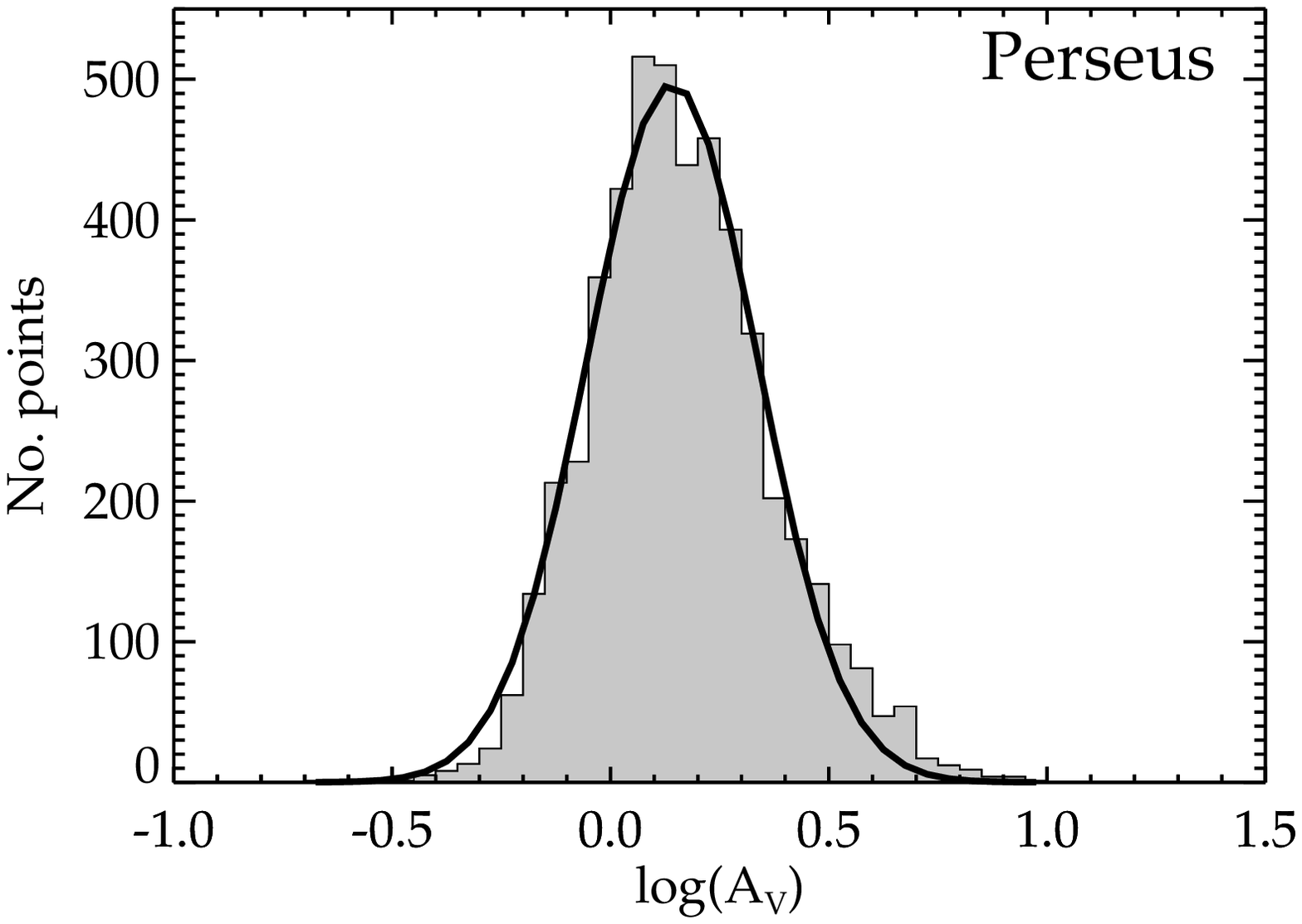}
\caption{
Histograms of extinction as derived from IRAS 60 and 100\,\um\
emission in Ophiuchus (left) and Perseus (right) for the areas shown
in figures
\ref{oph_2mass} and \ref{per_2mass} (filled histograms) overplotted
with a log-normal fit to the data (thick black curves; in the case of
Ophiuchus, a two component fit was used).
\label{iras_hist}}
\end{figure}

\begin{figure}
\plottwo{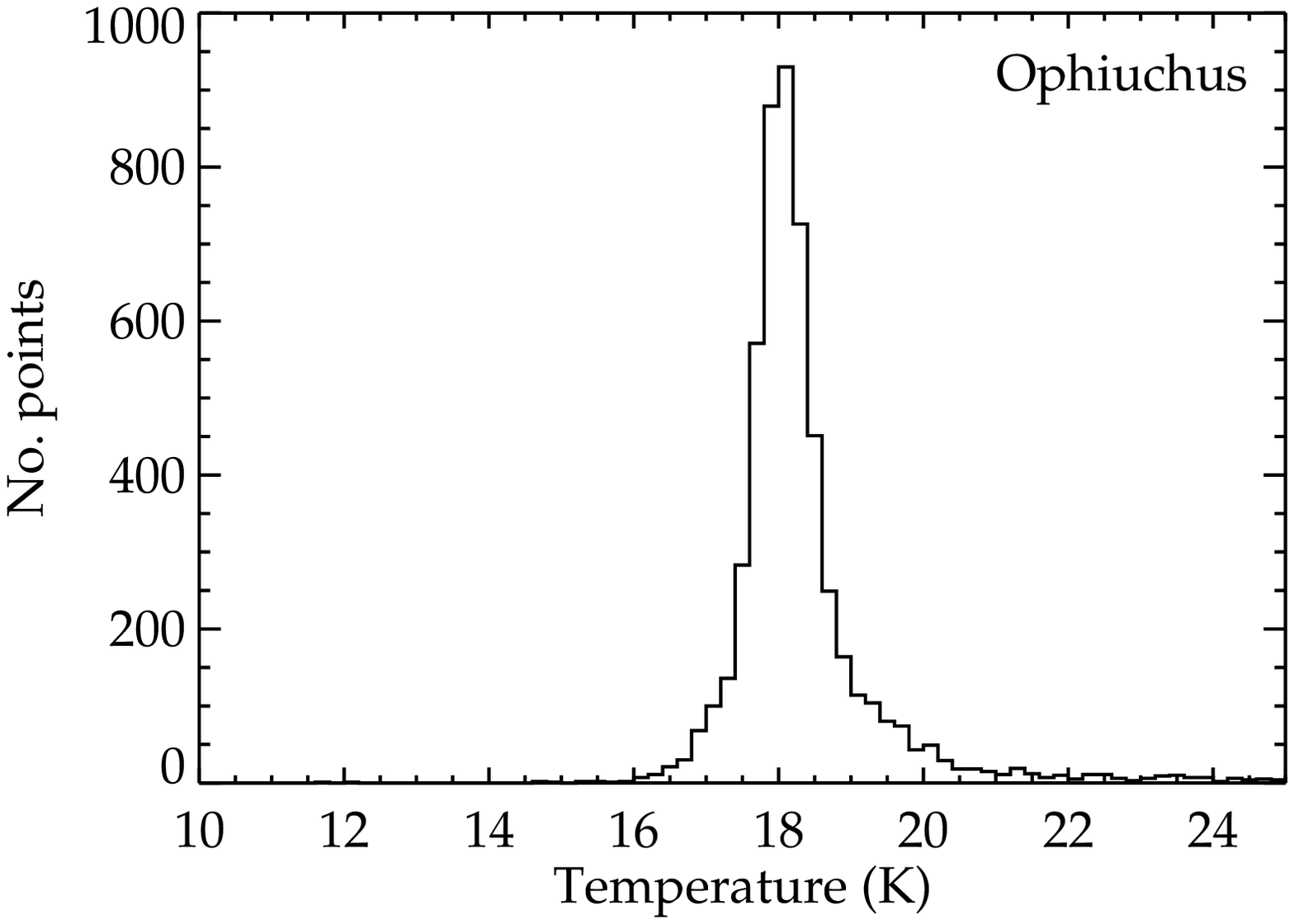}{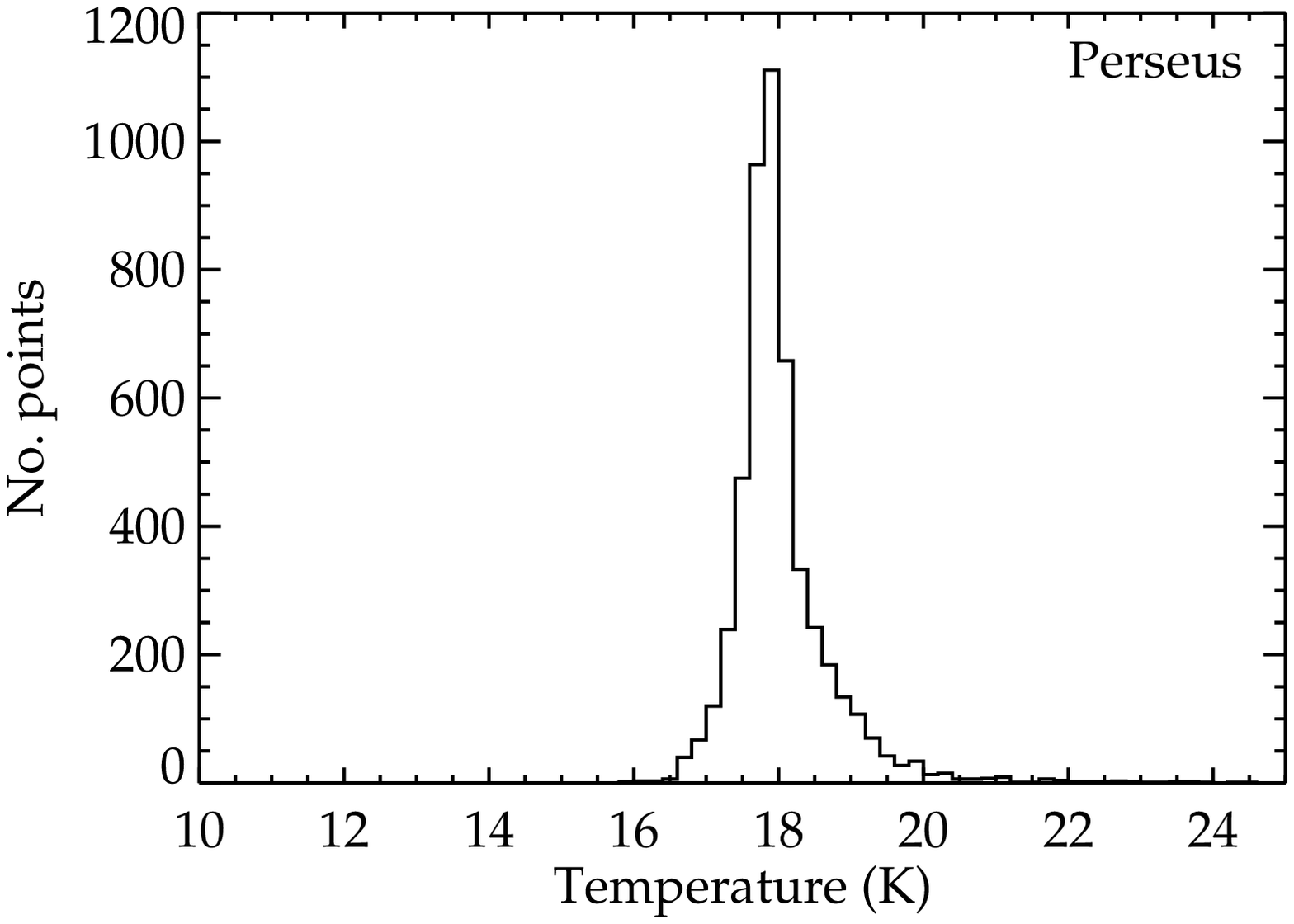}
\caption{Histograms of temperature as derived from IRAS
in Ophiuchus (left) and Perseus (right) for the areas shown in figures
\ref{oph_2mass} and \ref{per_2mass}. Note that the relatively low resolution
of these maps ($\sim 5'$) means that we are not sensitive to the
coldest densest regions within the clouds, and hence the average
temperature is somewhat higher than the $\sim$10\,K more typically
associated with molecular clouds.
\label{iras_temphist}}
\end{figure}

Histograms showing the distribution of extinctions and temperatures in
the two regions are shown in figures \ref{iras_hist} and
\ref{iras_temphist}.  Detailed discussion of the histograms and a comparison 
of the histograms produced from IRAS and 2MASS/NICER appear in
\citet{schnee_iras} and \citet{gsr} respectively, 
but they are included here for completeness. Note that due to their
differing resolution the histograms presented here are not directly
comparable with those presented for the 2MASS/NICER extinction and CO
data in sections \ref{nicer} and \ref{fcrao}.

\clearpage

\clearpage

\subsection{Atomic Hydrogen Maps from GBT}
\label{gbt}

Maps of 21\,cm HI emission over the Ophiuchus and Perseus clouds,
covering 5\,deg$^{2}$ and 20 \,deg$^{2}$ respectively, were obtained
with the 100\,m NRAO Green Bank Telescope\footnote{NRAO (and the GBT)
are operated by Associated Universities, Inc., under cooperative
agreement with the NSF.} (GBT) in West Virginia, USA over two
observing runs in 2004 March and 2005 April.  Both maps cover the
densest cores in the regions as well as regions of substantially lower
density, where HI could be stronger in emission due to the prevalence
of atomic hydrogen, rather than molecular hydrogen.  On-the-fly
mapping and frequency switching with a 1\,MHz throw were used together
with a data dumping rate of twice the Nyquist sampling rate, i.e.\ 4
dumps as the telescope moves over a whole beam.  The 12.5\,MHz total
bandwidth mode of the GBT Spectrometer was used with two spectral
windows, one at 1420.4\,MHz for HI, the other centered at 1666.4\,MHz
for the two OH lambda-doubling lines at 1667.4\,MHz and 1666.4
\,MHz\footnote{The OH data will be presented in a future paper and is
not discussed further here.}.  Each spectral window has two linear
polarizations, for a total of 4 IF inputs.  With 16,384 lags, a
0.76\,kHz channel width was achieved.

During reduction, frequency-switched data at both frequencies ($+/-$
frequency throw) were treated independently because of instrument
baseline stability.  The data were reduced using IDL routines written
by G. Langston that are consistent in terms of calibration when
checked with AIPS++ packages provided by NRAO.  Conversion to antenna
temperature was achieved through scaling by a noise tube input, and
conversion to absolute flux levels was achieved through comparisons
with observations of Mars.  Calibrated data were regridded to 4$'$
spacing in AIPS.  The final data have an angular resolution of
$\sim$9$'$ FWHM, a spectral resolution of 0.32\,\kms, and a typical
1-$\sigma$ rms noise of 0.15\,K per channel.  The properties of the HI
maps are summarised in table \ref{h1_data}.  3-D fits files can be
viewed online at the COMPLETE website -- the value of these data is in
the spectral information and hence we do not show an integrated
intensity map here.  The angular resolution is comparable to the size
of the dense structures in the two regions surveyed.
\begin{deluxetable}{ccc}
\tablewidth{0pc}
\tablecaption{HI Data
\label{h1_data}}
\tablehead{\colhead{}&\colhead{Ophiuchus}&\colhead{Perseus}}
\startdata
Pixel Size ($'$)&4&4\\
HPBW ($'$)&9&9\\
Areal Coverage (sq. degrees)& 5 & 20\\
1$\sigma$ rms/channel (K) &0.15 &0.15\\
Spectral Resolution (\kms) & 0.32 & 0.32
\enddata
\end{deluxetable}

The line profiles of HI in Ophiuchus reveal a strong and extensive HI
Narrow Self-Absorption \citep[HINSA; ][]{hinsa} component, which is
well correlated with molecular emission. This is consistent with
previous pointed observations of the same region at a much lower
resolution \citep{gh94}.  Channel maps of HI emission between 8 to
11\,\kms\ suggest a possible association of HI gas in this velocity
range with the heated ring seen in warm the IRAS temperature map.

\begin{figure}
\plotone{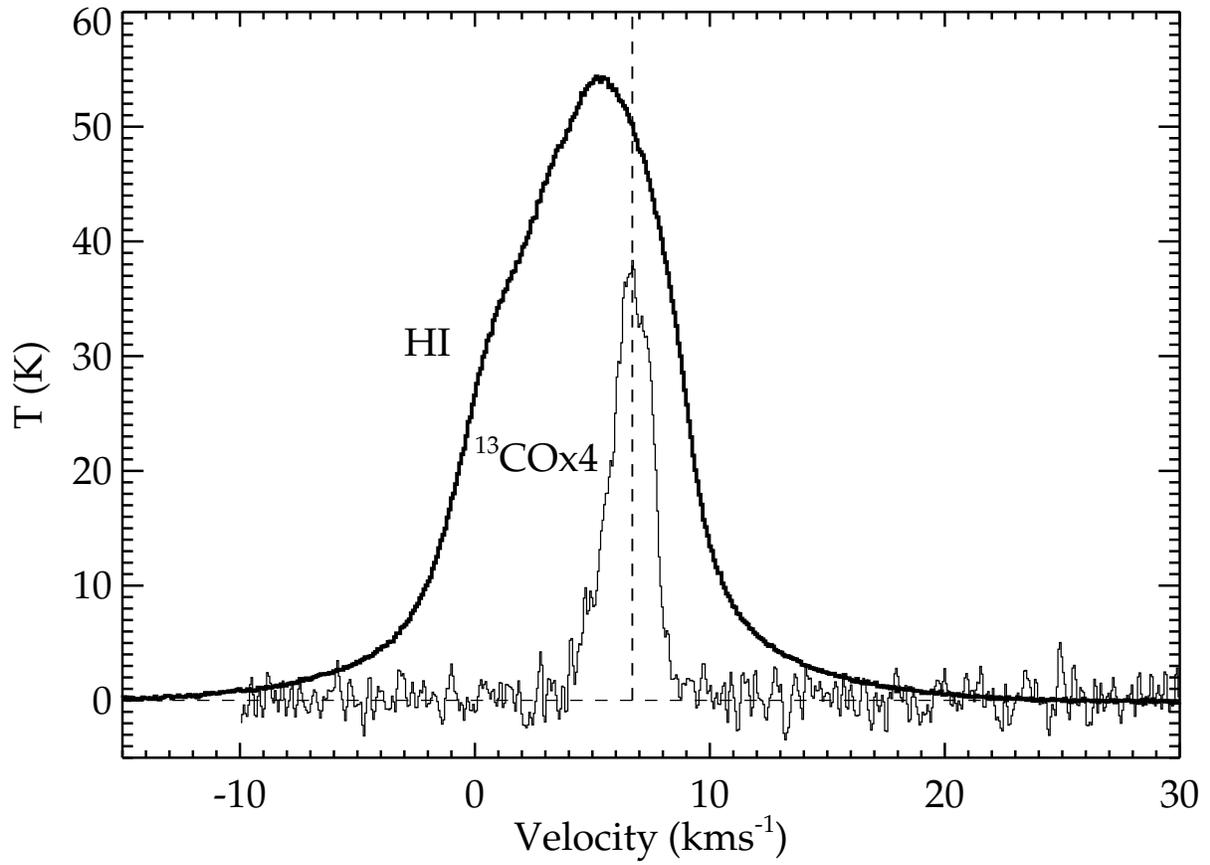}
\caption{Spectra of HI and \xco\ at the position of the dense core B1. 
The molecular gas (as traced by the \xco) has a peak velocity offset
by $\sim$2\,\kms\ from the atomic gas.  Both spectra are centered on
the position 03$^{\rm h}$33$^{\rm m}$17\fs8 +31\degrees07$^{\rm
m}$30$^{\rm s}$ (J2000).
\label{b1_h1}}
\end{figure}

The main component of HI emission toward the line of sight of Perseus
is centered around 4 to 8\,\kms, with the velocity of peak emission
becoming redder toward the west of the region, as is seen in the
molecular gas. The HI peaks, however, tend to be on the bluer side of
the molecular gas by 1--2\,\kms. For example, around dense core B1,
the HI emission shows approximately a single Gaussian profile with
peak velocity at 4.8\,\kms, while \xco\ peaks at $\sim$6.7\,\kms\
(figure \ref{b1_h1}).

\clearpage

\clearpage

\begin{figure}
\plotone{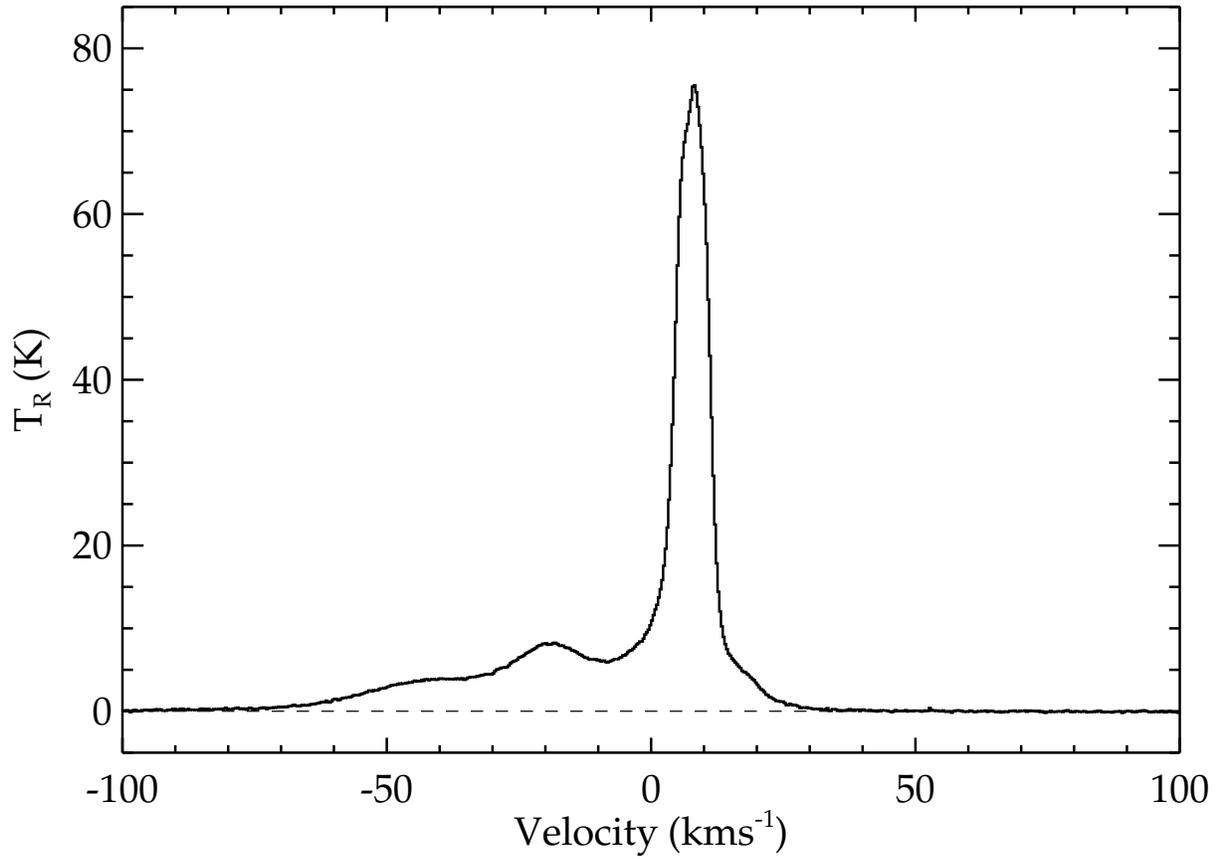}
\caption{Spectrum of HI emission at the position 03$^{\rm h}$46$^{\rm m}$21\fs6 +33\degrees26$^{\rm m}$48\fs5 (J2000), showing an extended line wing and second component, possibly from another Galactic arm.
\label{wing}}
\end{figure}

The line-width of HI emission in Perseus is around 7--10\,\kms\ FWHM.
Unlike other nearby regions, such as Taurus and Ophiuchus, the HINSA
component is only seen toward a small portion of the dense
clouds. This may be explained by viewing geometry and the different
distances of the Perseus components.  A minor, but very interesting
component in our Perseus map is the presence of high velocity line
wings in HI emission, extending from the main component all the way
down to $-$50\,\kms, where it peaks up again to possibly reflect HI
emission from another galactic arm (figure \ref{wing}). Although its
origin is unclear at the moment, it is worth noting that it is
probably too wide to be explained by a single low level HI emission
component somewhere else in the galaxy.

\clearpage

\clearpage

\subsection{Molecular Line Maps from FCRAO}
\label{fcrao}

Observations in the \co\ 1--0 (115.271\,GHz) and \xco\ 1--0
(110.201\,GHz) transitions were carried out throughout the 2002--2005
observing seasons at the 14\,m Five College Radio Astronomy
Observatory\footnote{ FCRAO is supported by NSF Grant AST 02-28993.}
(FCRAO) telescope in New Salem, MA, U.S.A.  The SEQUOIA 32-element
focal-plane array and an On-the-Fly mapping technique were used to
make $10'\times 10'$ submaps.  The dual-IF narrowband digital
correlator enabled \co\ and \xco\ to be observed simultaneously.  The
correlator was used in a mode that provided a total bandwidth of
25\,MHz with 1024 channels in each IF, yielding an effective velocity
resolution of 0.07\,\kms.  Data were taken during a wide range of
weather qualities and system temperatures were generally between 500
and 1000\,K at 115\,GHz and 200 and 600\,K at 110\,GHz (single
sideband).  Due to its low elevation, system temperatures for
Ophiuchus were consistently higher than for Perseus.  Submaps with
higher system temperatures were repeated to achieve uniform
sensitivity where possible.

The submaps were obtained by scanning in the Right Ascension
direction\footnote{A rotation angle of 326 degrees east of north was
used as the scanning direction in the case of Perseus.}, and an
off-source reference scan was obtained after every two or four rows,
depending on weather and elevation.  Off-positions were checked to be
free of emission by making separate 10$'$ OTF maps, with an
off-position an additional 30$'$ offset.  Calibration was found via
the chopper-wheel technique
\citep{ku81}, yielding spectra with units of ${\rm{T}_A}^*$. Pointing 
was checked regularly and found to vary by less than 5$''$ rms.  

\begin{figure}
\plottwo{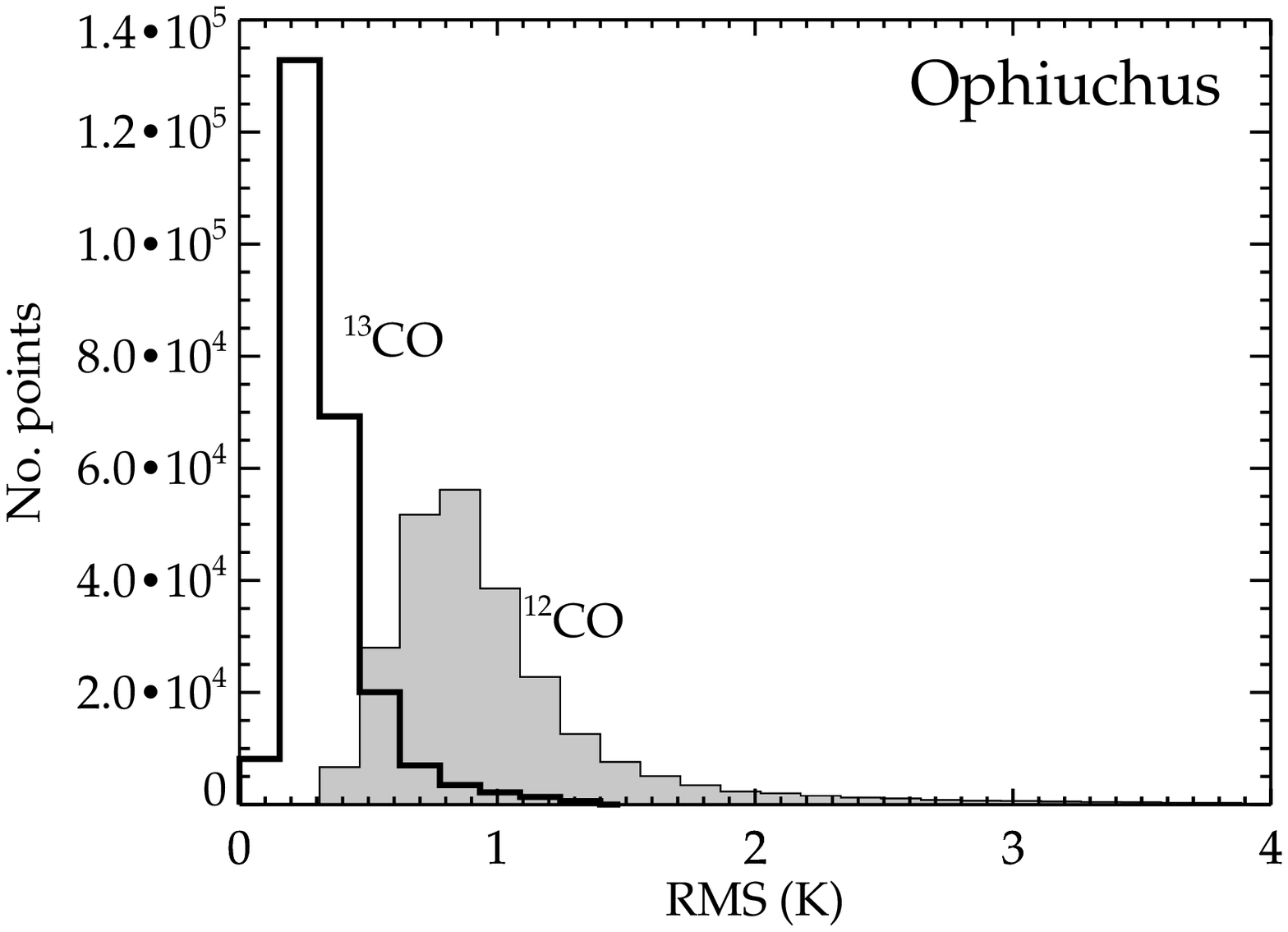}{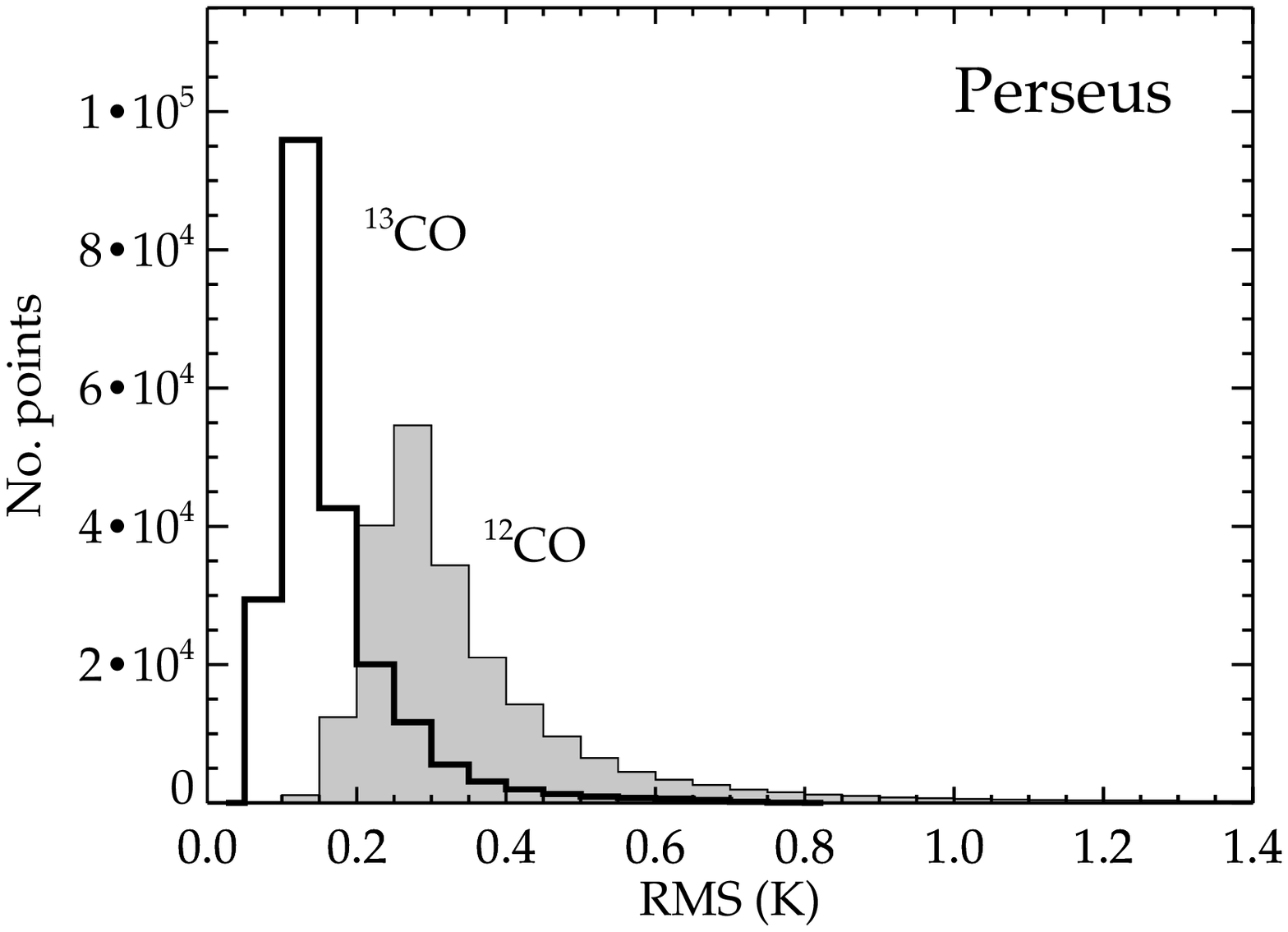}
\caption{Histograms of rms noise per channel in the \co\ and \xco\ 
maps of Ophiuchus (left) and Perseus (right).
\label{rms}}
\end{figure}

Due to dewar rotation, the OTF data are not not evenly sampled, and so
a convolution and regridding algorithm has to be applied to the data
to obtain spectra on a regularly sampled grid.  This process was
carried out on the individual $10'\times 10'$ submaps using software
provided by the observatory \citep{hnb01}.  After subtraction of a
linear baseline, each spectrum was convolved with nearby spectra onto
a regular 23$''$ grid weighted by rms$^{-2}$, yielding a
Nyquist-sampled map.  The submaps were then combined into the final
map using an IDL routine, and corrected for the main beam efficiency
(0.5 and 0.45 at 110 and 115\,GHz respectively). Due to the nature of
the OTF technique, spectra in pixels near the edges of the map have a
significantly higher rms noise. Hence an rms filter was applied to the
combined map to blank pixels which had an rms noise of more than 3
times the mean rms noise for the entire map. Histograms, showing the
range of rms noise in the final 3-dimensional data cubes are presented
in figure
\ref{rms}.

\clearpage

\clearpage

\begin{figure}
\plottwo{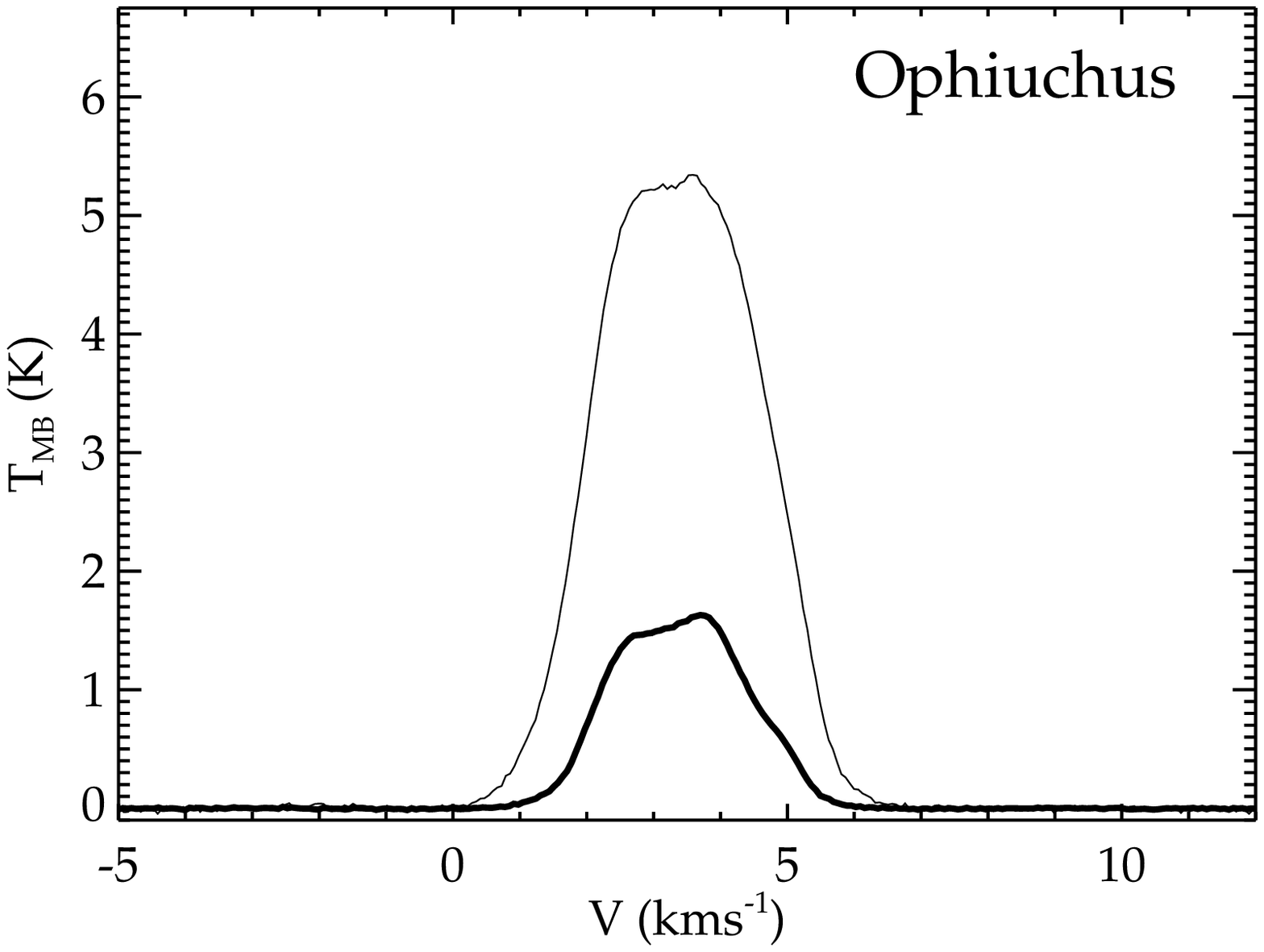}{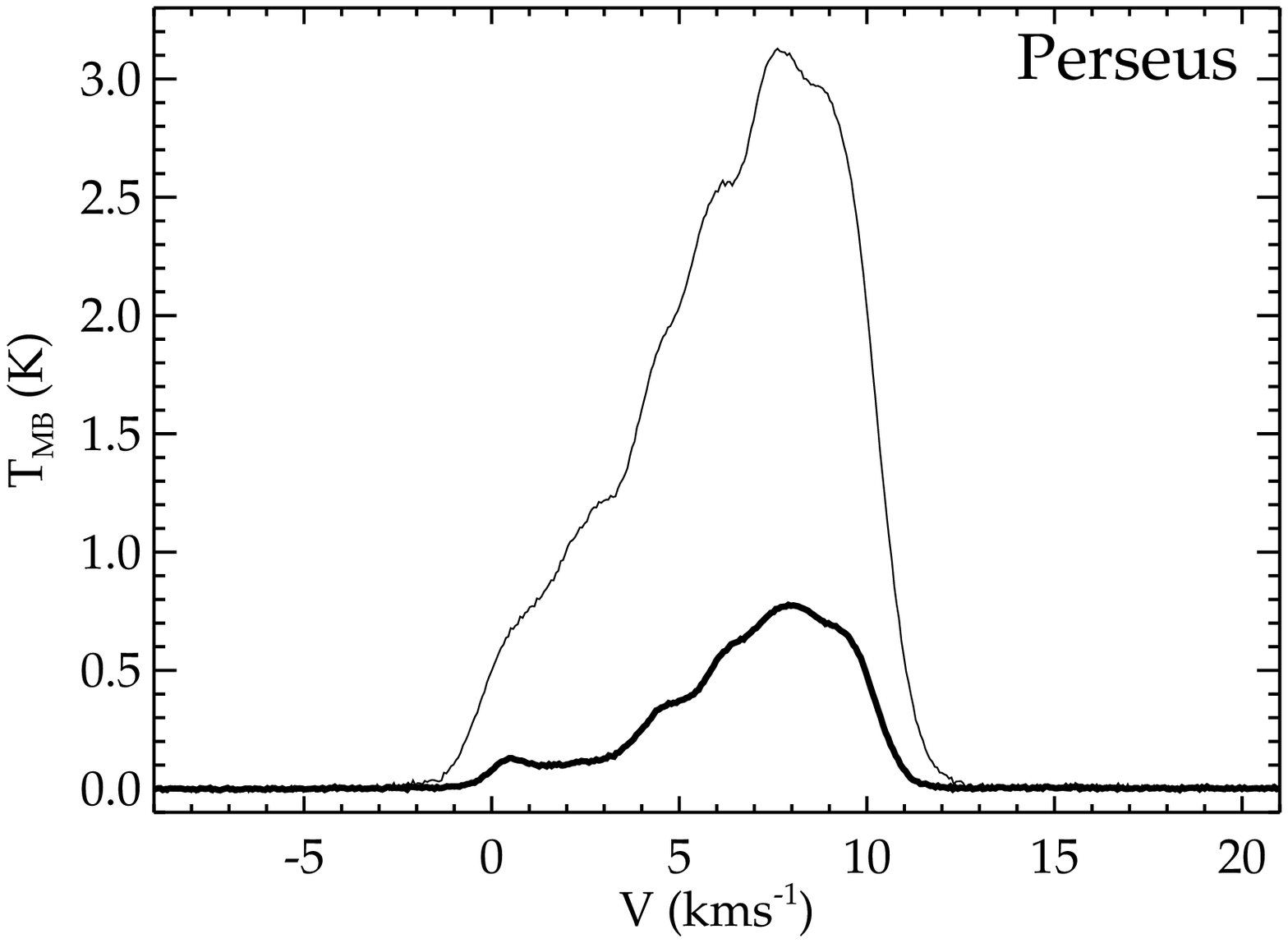}
\caption{Average \co\ (thin lines) and \xco\ (thick lines) spectra 
for Ophiuchus (left) and Perseus (right), created by summing the
spectra in all pixels where the ratio of peak antenna temperature to
rms noise is was greater than 3. The multi-component nature of Perseus
is clearly visible, whilst Ophiuchus displays a more Gaussian-like
profile.
\label{spec}}
\end{figure}

The resolution, coverage, sampling and sensitivity of the resulting
data cubes are summarised in table \ref{co_data}. Figure \ref{spec}
shows average \co\ and \xco\ spectra for Ophiuchus and Perseus, created
by summing the spectra in all pixels where the ratio of peak antenna
temperature to rms noise is greater than 3. 
\begin{deluxetable}{cccc}
\tablewidth{0pc}
\tablecaption{CO Data
\label{co_data}}
\tablehead{\colhead{}&\colhead{}&\colhead{Ophiuchus}&\colhead{Perseus}}
\startdata
&Pixel Size ($''$)&23&23\\
&Total number of spectra &244874 & 214316\\
&Areal Coverage (sq. degrees)& 10.0 & 8.7\\
\\
\tableline
\\
&HPBW ($''$)&46&46\\
{\bf \co}&Mean RMS/channel\tablenotemark{1} (K)&0.98 &0.35\\
&Min RMS/channel (K) &0.31 &0.11\\
&Spectral Resolution (\kms) &0.064 & 0.064\\
\\
\tableline
\\
&HPBW ($''$)&44&44\\
{\bf \xco}&Mean RMS/channel\tablenotemark{1} (K)&0.33&0.17 \\
&Min RMS/channel (K) &0.10&0.06\\
&Spectral Resolution (\kms)& 0.066 & 0.066\\
\enddata
\tablenotetext{1}{After blanking of pixels with an rms noise greater than 3 times the mean rms in the unblanked data file.}
\end{deluxetable}

\clearpage

\clearpage

Integrated intensity maps of \co\ and \xco\ emission were made by
summing over the range of velocities in which emission was seen. The
integrated intensity maps of \xco\ in Ophiuchus and Persus are shown
in figures \ref{oph_co_scuba} and \ref{per_co_scuba}. The 
\co\ maps are not shown here but can be viewed online at
the COMPLETE website.

\clearpage
\begin{figure}
\caption{Integrated intensity image of \xco\ emission in Ophiuchus, overlaid
with the positions of the dense cores detected in submillimeter
continuum emission (red circles; see section
\ref{scuba}). Symbol size is proportional to the mass of the 
core.  The black contour indicates A$_V$=3 mag.\ from the 2MASS/NICER
map, and the grey contour indicates A$_V$=15 mag., the implied
threshold for dense core formation in Ophiuchus.
\label{oph_co_scuba}}
\end{figure}

\clearpage
\begin{figure}
\caption{Integrated intensity image of \xco\ emission in Perseus, overlaid
with the positions of the dense cores detected in submillimeter
continuum emission (red circles; see section
\ref{scuba}). Symbol size is proportional to the mass of the 
core.  The black contour indicates A$_V$=3 mag.\ from the 2MASS/NICER
map, and the grey contour indicates A$_V$=6 mag., the implied
threshold for dense core formation in Perseus.
\label{per_co_scuba}}
\end{figure}

Although extensive, the CO maps are more limited in areal coverage
than the preceding datasets. For instance, the northeast filament and
north west extension we see in the extinction map of Ophiuchus are not
well sampled in CO emission.  While the morphology generally follows
that of the extinction map, a bright maxima north of the Oph A core
within L1688 is quite prominent in the
\xco\ map of Ophiuchus. The average \co\ and
\xco\ spectra (figure \ref{spec}, left panel) show an approximately
Gaussian profile with a maximum width of $\sim$7\,\kms. Table
\ref{gaussfit} gives the central velocity and FWHM obtained by
fitting a Gaussian profile to the average spectra.
\begin{deluxetable}{ccc}
\tablewidth{0pc}
\tablecaption{Results of Gaussian fits to the average \co\ and 
\xco\ lines in Ophiuchus.
\label{gaussfit}}
\tablehead{\colhead{}&\colhead{FWHM}&\colhead{V$_{\rm LSR}$}\\
\colhead{}&\colhead{\kms}&\colhead{\kms}}
\startdata
\co & 2.77 & 3.38\\
\xco & 2.38 & 3.38
\enddata
\end{deluxetable}

The morphology of the integrated CO intensity in Perseus is again
similar to that of the extinction, but due to their $\sim$4 times
better linear spatial resolution over the extinction map, the \co\ and
\xco\ maps reveal complex substructure within the clumps we see in
extinction. The CO emission shows components at multiple velocities,
and a steep velocity gradient across the cloud complex, with a
difference of almost 10\,\kms\ over the $\sim$30\,pc east-west extent
of the complex. This is the main cause of the wide line-widths ($\sim$
15\,\kms) and non-Gaussian profiles exhibited by the average
\co\ and
\xco\ spectra shown in figure \ref{spec}\footnote{Although multiple
outflows in the region also contribute.}, and suggests that the
Perseus complex is much more dynamic than Ophiuchus.

\begin{figure}
\plottwo{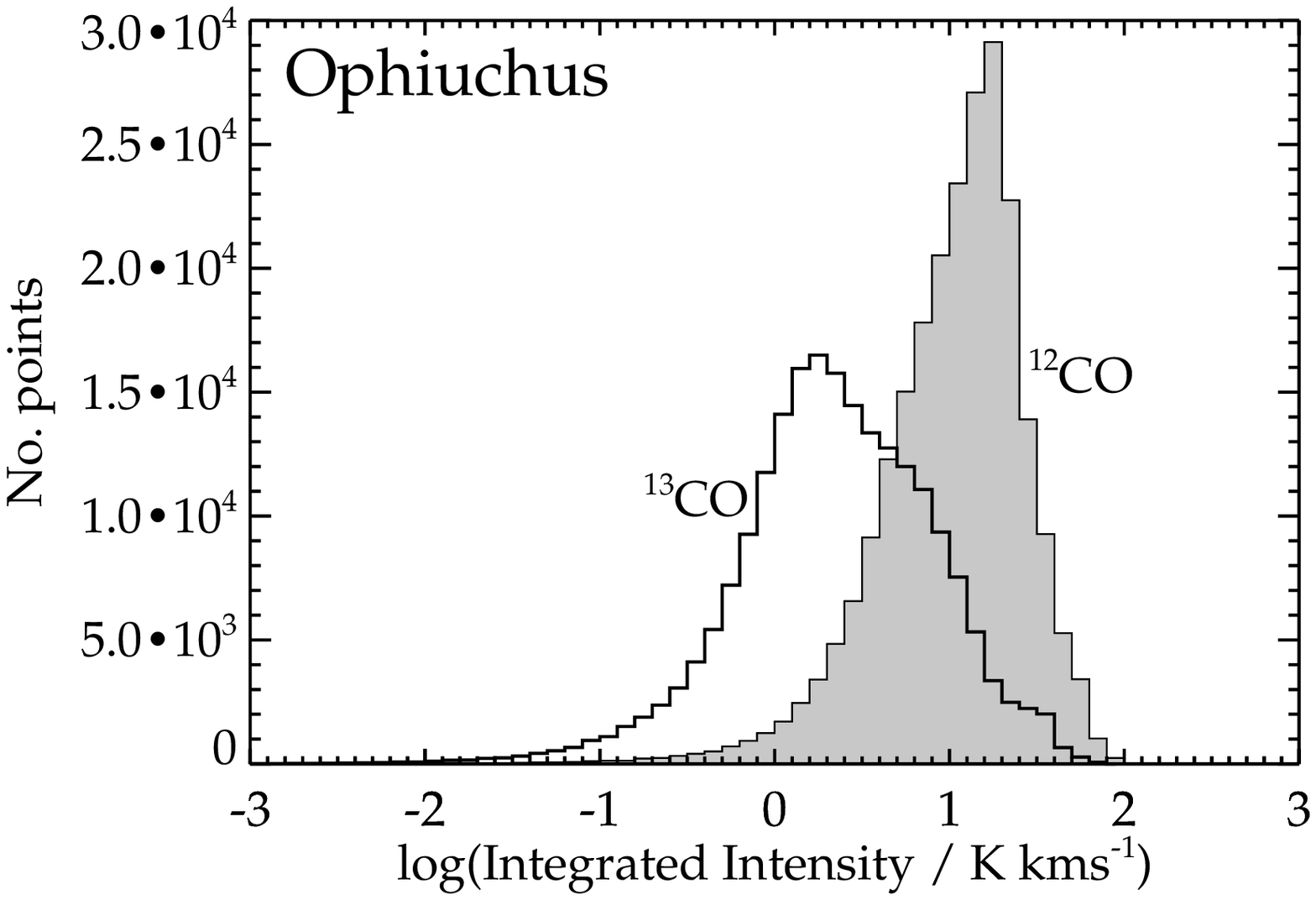}{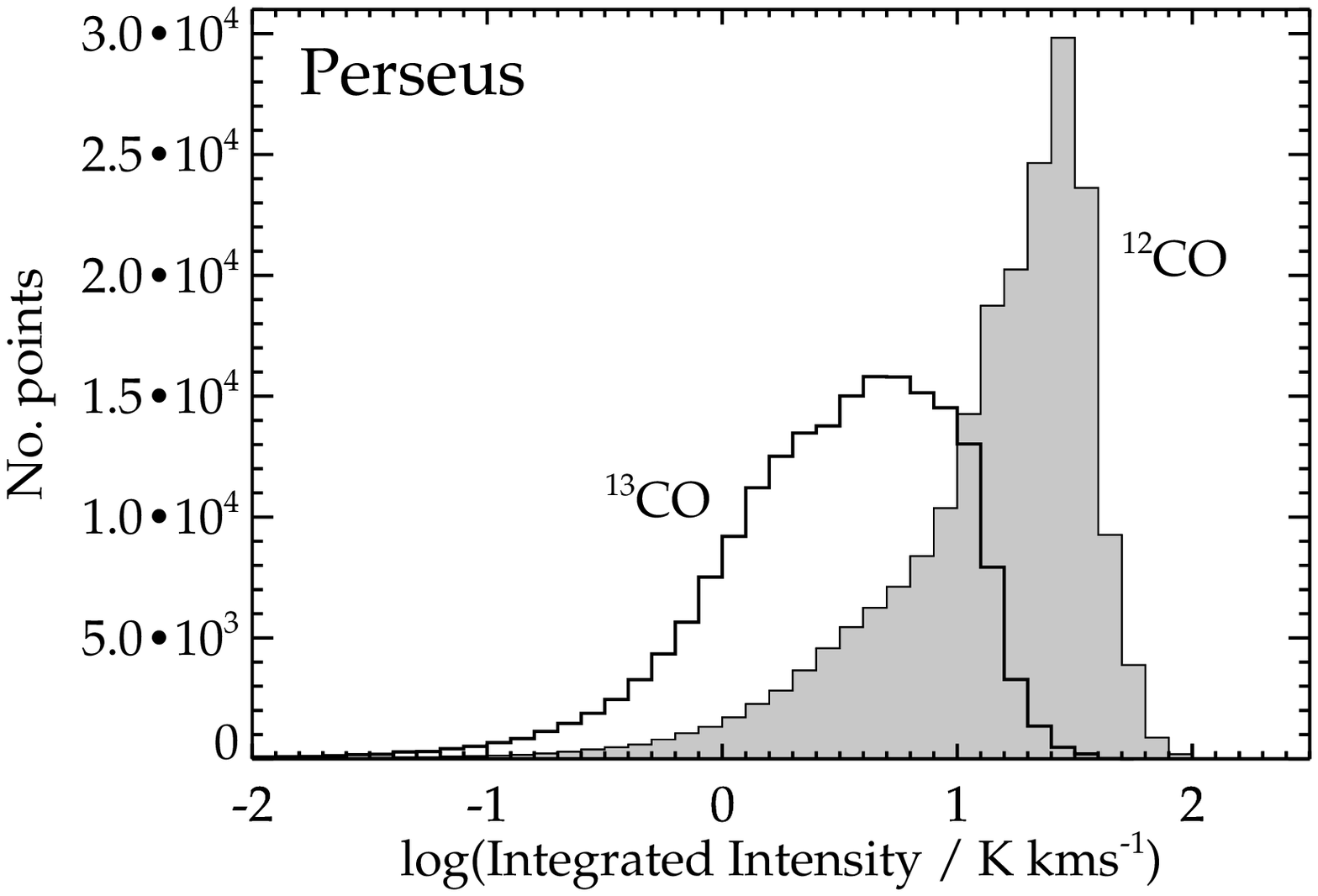}
\caption{Histograms of \co\ and \xco\ integrated
intensity for Ophiuchus (left) and Perseus (right).
\label{co_hist}}
\end{figure}

Histograms of the \co\ and \xco\ integrated intensity are shown in
figure \ref{co_hist}. Unlike the 2MASS/NICER extinction maps, the
distribution of CO intensities does not follow a log-normal
distribution. In particular there is a significant low-intensity tail
to the distribution. In Perseus, the distribution also appears
somewhat truncated on the high-intensity side. This is likely a result
of a combination of chemical effects (e.g.\ freeze-out) and high
optical-depth of both \co\ and \xco\ at higher column densities. The
warmer temperatures (as traced by the IRAS temperature map) in the
densest regions of Ophiuchus may prevent CO freezing out there.

\clearpage

\clearpage

\subsection{Thermal Dust Emission Maps from JCMT}
\label{scuba}

Submillimeter continuum data at 850\,$\mu$m of the Ophiuchus and Perseus 
molecular clouds were obtained using the Submillimetre Common User Bolometer 
Array (SCUBA) on the 15\,m James Clerk Maxwell Telescope \footnote{The JCMT 
is operated by the Joint Astronomy Centre in Hilo, Hawaii on behalf of the 
parent organizations Particle Physics and Astronomy Research Council in 
the United Kingdon, the National Research Council of Canada and The 
Netherlands Organization for Scientific Research.} (JCMT) on Mauna Kea, 
HI, U.S.A.  The data presented here are a combination of our own 
observations ($\sim$5.6\,deg$^{2}$ of Ophiuchus and $\sim$1.3\,deg$^{2}$ 
of Perseus) with publicly available archival data\footnote{Guest User, 
Canadian Astronomy Data Centre, which is operated by the Dominion 
Astrophysical Observatory for the National Research Council of Canada's 
Herzberg Institute of Astrophysics.} for a total of $\sim$5.8 deg$^{2}$ 
of Ophiuchus and $\sim$3.5\,deg$^{2}$ of Perseus.  

All raw data were first flat-fielded and extinction corrected using
the standard SCUBA software \citep{holland99}.  The data were then
converted into images by applying the matrix inversion technique of
\citet{johnstone00a} with a pixel size of 6\arcsec.  Although the maps
have an intrinsic beamsize of 14\arcsec, their effective beamsize is
19.9\arcsec\ because each was convolved with a $\sigma$ = 6\arcsec\
Gaussian to reduce pixel noise.  Structure on scales several times
larger than the chop throw ($>$120\arcsec) may be an artifact of image
reconstruction (independent of the technique; see
\citealt{johnstone00a}).  Large-scale artifacts were removed by
subtracting a convolved version of the original map (made with a
Gaussian of $\sigma$ = 90\arcsec) from each map.  To minimize the
occurrence of artificial negative ``bowls" around bright sources
resulting from this technique, all pixels with values $>\vert5\vert$
times the mean noise were set to specifically +5 or -5 times the mean
noise prior to convolution.  Since each map was constructed from data
obtained under a variety of weather conditions, the noise across each
map is not uniform.  The noise variation across each map, however, is
typically only a factor of a few.  The mean and rms standard deviation
are $\sim$40 and 20\,mJy beam$^{-1}$ respectively in Ophiuchus and
$\sim$8 and $\sim$7\,mJy beam$^{-1}$ respectively in Perseus. Due to
their size, it is not possible to display the maps clearly in this
work, and again we refer the reader to the COMPLETE website where fits
files are publically available. The properties of the maps are
summarised in table \ref{scuba_data}.
\begin{deluxetable}{ccc}
\tablewidth{0pc}
\tablecaption{SCUBA 850\um\ Continuum Data
\label{scuba_data}}
\tablehead{\colhead{}&\colhead{Ophiuchus}&\colhead{Perseus}}
\startdata
Pixel Size ($''$) &6 &6 \\
Effective Resolution ($''$) & 19.9 & 19.9\\
Areal Coverage (sq. degrees) & 5.8 & 3.5\\
Mean RMS noise\tablenotemark{1} (mJy\,beam$^{-1}$)& 40 (20) & 8 (7)\\
\enddata
\tablenotetext{1}{Values in brackets are RMS standard deviation.}
\end{deluxetable}

Detailed analyses and images of the submillimeter maps are presented
in \citet*{johnstone04} for Ophiuchus and
\citet*{helen} for Perseus, or can be viewed online
at the COMPLETE website so are not repeated here. The positions of the
dense clumps detected in the submillimeter are shown as red circles in
figures \ref{oph_co_scuba} and \ref{per_co_scuba}.

Of particular interest in the Ophiuchus map is the lack of any
submillimeter emission northwest of L1688, which was well covered with
SCUBA as a result of the IRAS-based extinction map showing possible
moderate extinction in that direction.  This lack of detection is
likely due to especially low sensitivity to extended, diffuse
structures that may inhabit that location (see
\citealt{johnstone04}).  Also, as has
been discussed in \citet{johnstone04}, there appears to be threshold
of column-density (extinction) below which no dense cores are
found. In the case of Ophiuchus this threshold is at an A$_V$ of
$\sim$15\,mag, as indicated by the grey contour in figure
\ref{oph_co_scuba}.

Like in Ophiuchus, all of the dense submillimeter clumps in Perseus
are found to lie within the larger high-extinction regions, but the
threshold of $\sim$5--7\,A$_V$ is somewhat lower than in
Ophiuchus. This difference in thresholds cannot be attributed to the
different sensitivities of the two submillimeter maps -- this is
discussed further in
\citet{helen}. A similar threshold has also been reported by
\cite{hatchell05} and \citet{enoch}.

\clearpage
\section{Summary}
We have presented maps of the gas and dust in the Ophiuchus and
Perseus star-forming regions, obtained using a range of different
techniques, and providing information on the star-forming material on
scales of ~0.1-10pc. These observations complement the observations
made as part of the Spitzer Space Telescope ``From Molecular Cores to
Planet Forming Disks'' Legacy Program, which provides a catalogue of
the young stars.

Some highlights of the data are:

\begin{itemize}
\item The apparent morphology and
distribution of star-forming material varies significantly between
different column-density tracers, with the near-infrared extinction
method providing the closest match to the log-normal distribution of
material predicted by numerical simulations.  

\item CO emission in Ophiuchus
shows approximately Gaussian line shapes, while Perseus appears much
more dynamic, with multiply peaked, wide, non-Gaussian lines.

\item There appears to be an extinction threshold below which dense sub-mm
cores are not detected. This threshold varies from region to region
(A$_V\sim$15 in Ophiuchus and $\sim$5-7 in Perseus).

\item HI emission in Perseus shows an extremely high-velocity line wing, 
possibly reflecting emission from another Galactic arm.

\end{itemize}

All of the data sets presented here are publically available from the
COMPLETE website.

\acknowledgments
NAR is supported by the National Science Foundation through award
AST-0407172. JBF is supported by the NASA ADP program.  HK is
supported by a National Research Council of Canada GSSSP award. DJ's
research is supported by a grant from the Natural Sciences and
Engineering Research Council of Canada. JEP is supported by the
National Science Foundation through grant AF002 from the Association
of Universities for Research in Astronomy, Inc., under NSF cooperative
agreement AST-9613615 and by Fundaci\'on Andes under project
no. C-13442. SLS is supported by a National Science Foundation
Graduate Research Fellowship.

{\it Facilities:} \facility{FCRAO}, \facility{JCMT},
\facility{IRAS}, \facility{GBT}, \facility{2MASS}.





\bibliographystyle{apj}
\bibliography{refs}

\end{document}